\documentclass[lettersize,journal]{IEEEtran}
\usepackage{amsmath,amsfonts}
\usepackage{algorithmic}
\usepackage{array}
\usepackage[caption=false,font=normalsize,labelfont=sf,textfont=sf]{subfig}
\usepackage{textcomp}
\usepackage{stfloats}
\usepackage{url}
\usepackage{verbatim}
\usepackage{graphicx}
\def\BibTeX{{\rm B\kern-.05em{\sc i\kern-.025em b}\kern-.08em
    T\kern-.1667em\lower.7ex\hbox{E}\kern-.125emX}}
\usepackage{balance}

\usepackage{threeparttable}
\usepackage{enumitem}            
\usepackage{amssymb}            
\usepackage{multirow}            
\usepackage{booktabs}            
\usepackage[ruled]{algorithm2e}  
\usepackage{caption}            
\usepackage{array}
\usepackage[caption=false,font=normalsize,labelfont=sf,textfont=sf]{subfig}

\begin{document}
\title{From Classification to Consistent Templates: Multiple Permuted-Label Classifier Encoding for Biometric Template Protection}
\author{\IEEEauthorblockN{Baogang Song\IEEEauthorrefmark{1},
Zhongshu Zhao\IEEEauthorrefmark{1}, 
Qianrong Zheng\IEEEauthorrefmark{1},
Jianwen Xiang\IEEEauthorrefmark{1}} and
Dongdong Zhao\thanks{This work was supported by the Fundamental Research Funds for the Central Universities under Grant 104972026ZHZXhp0024. (Corresponding author: Dongdong Zhao, e-mail: zdd@whut.edu.cn.)}\IEEEauthorrefmark{1}

\IEEEauthorblockA{\IEEEauthorrefmark{1}School of Artificial Intelligence, Wuhan University of Technology}}

\markboth{Journal of \LaTeX\ Class Files,~Vol.~18, No.~9, September~2020}%
{How to Use the IEEEtran \LaTeX \ Templates}

\maketitle

\begin{abstract}
Biometric template protection (BTP) must secure stored templates while tolerating intra-class variations. Existing methods rely on protected-domain similarity matching, error correction, or predefined-template mappings, potentially retaining exploitable similarity structures, introducing helper-data risks, depending on artificial targets, or coupling protection to specific modalities. Storing only cryptographic hash digests eliminates directly comparable representations and conceals pre-hash templates, but hash-based exact-match verification requires genuine samples to generate identical intermediate templates before hashing. Identity classification is naturally suited to this requirement because it maps variable biometric samples to stable and discriminative identity-level outputs. Based on this insight, we propose Multiple Permuted-Label Classifier Encoding (MPLCE). Through classifier-specific label permutations, MPLCE assigns each identity different labels across multiple classifiers. The predicted labels are encoded and concatenated to form an intermediate template, preventing repeated encodings of a single identity label and enlarging the effective candidate space while preserving classification consistency. The template is randomized with an application-specific XOR string and cryptographically hashed, enabling exact-match verification without error correction codes or biometric-dependent helper data. Using modality-specific classifiers, MPLCE retains the same template generation and protection procedure across modalities. On four face and two iris datasets, MPLCE achieves competitive performance, including a GAR of 98.61\% at a FAR of 5.51\(\times\)10\textsuperscript{-5}\% on YTF and a GAR of 99.10\% at a FAR of 0.00\% on CASIA-Iris-Lamp. Security analyses and attack evaluations support its irreversibility, revocability, and unlinkability under the threat model.

\end{abstract}


\section{Introduction}

Biometric recognition is increasingly deployed for identity authentication in access control, mobile payments, forensics, and border inspection~\cite{ref_biometric_recognition}. Compared with passwords and physical tokens, biometric traits provide convenience and strong identity binding. However, the long-term storage of biometric templates introduces serious security and privacy concerns~\cite{ref_biometric_template_security,ref_iso24745}. Once compromised, these templates may be exploited for identity impersonation and cross-application linkage, and may expose sensitive biometric information. More critically, biometric traits are persistent and non-replaceable; compromised biometric information cannot be reset or reissued in the same manner as passwords or tokens. Therefore, secure and privacy-preserving generation, storage, and verification of biometric templates has become a fundamental problem in BTP.

To mitigate these risks, numerous BTP methods have been proposed to support irreversibility, revocability, and unlinkability~\cite{ref_iso24745}. Traditional BTP methods are commonly divided into cancelable biometrics (CB) and biometric cryptosystems (BCS)~\cite{ref_biometric_template_security,ref4}. CB methods protect biometric templates through noninvertible and revocable transformations. However, many CB schemes perform similarity matching in the protected domain~\cite{ref9,ref12,ref15} and therefore retain matchability structures that may be exploited for similarity leakage or cross-application linkage~\cite{ref10,ref11,ref13,ref14,ref16}. BCS, including fuzzy commitment, fuzzy vault, and fuzzy extractor schemes, bind or derive cryptographic keys from noisy biometric data~\cite{ref19,ref17,ref21}. These schemes generally rely on error correction or recovery mechanisms and public helper data to tolerate intra-class variations, which may introduce additional security and privacy risks~\cite{ref23,ref24,ref25,ref26,ref27,ref29,ref51}. Moreover, many existing BTP methods are coupled with modality-specific feature representations, transformations, or matching rules, making it difficult to formulate a unified template generation principle across biometric modalities~\cite{ref_cb_benchmark}.

Existing neural network (NN)-based BTP studies use neural networks to learn protected representations, stable codes, or mappings from biometric samples to predefined user templates~\cite{ref4,ref34,ref35,ref_biodeephash,ref32,ref33,ref62}. Many of these methods tolerate residual intra-class variations through protected-domain similarity matching, secure sketches, or error correction mechanisms, and therefore retain comparable protected representations or rely on auxiliary recovery structures, which may introduce additional security and privacy risks. Hash-based methods such as BioDeepHash reduce this exposure by combining learned stable codes with application-specific randomization and cryptographic hashing so that only hash digests are stored~\cite{ref_biodeephash}. However, hash-based exact-match verification requires genuine samples from the same identity to generate exactly the same intermediate template before hashing. For existing NN-based methods that adopt cryptographic hashing, the network must directly generate the complete intermediate template, with all bit positions remaining identical across samples from the same identity. Achieving such exact consistency is difficult because biometric samples from the same subject naturally exhibit intra-class variations caused by pose, illumination, acquisition conditions, and other factors~\cite{ref_biometric_recognition}, while a difference at any bit position changes the hash input and leads to verification failure. Consequently, directly generating identity-discriminative and intra-class-consistent intermediate templates remains a key challenge.

\begin{figure*}
    \centering
    \includegraphics[width=\textwidth]{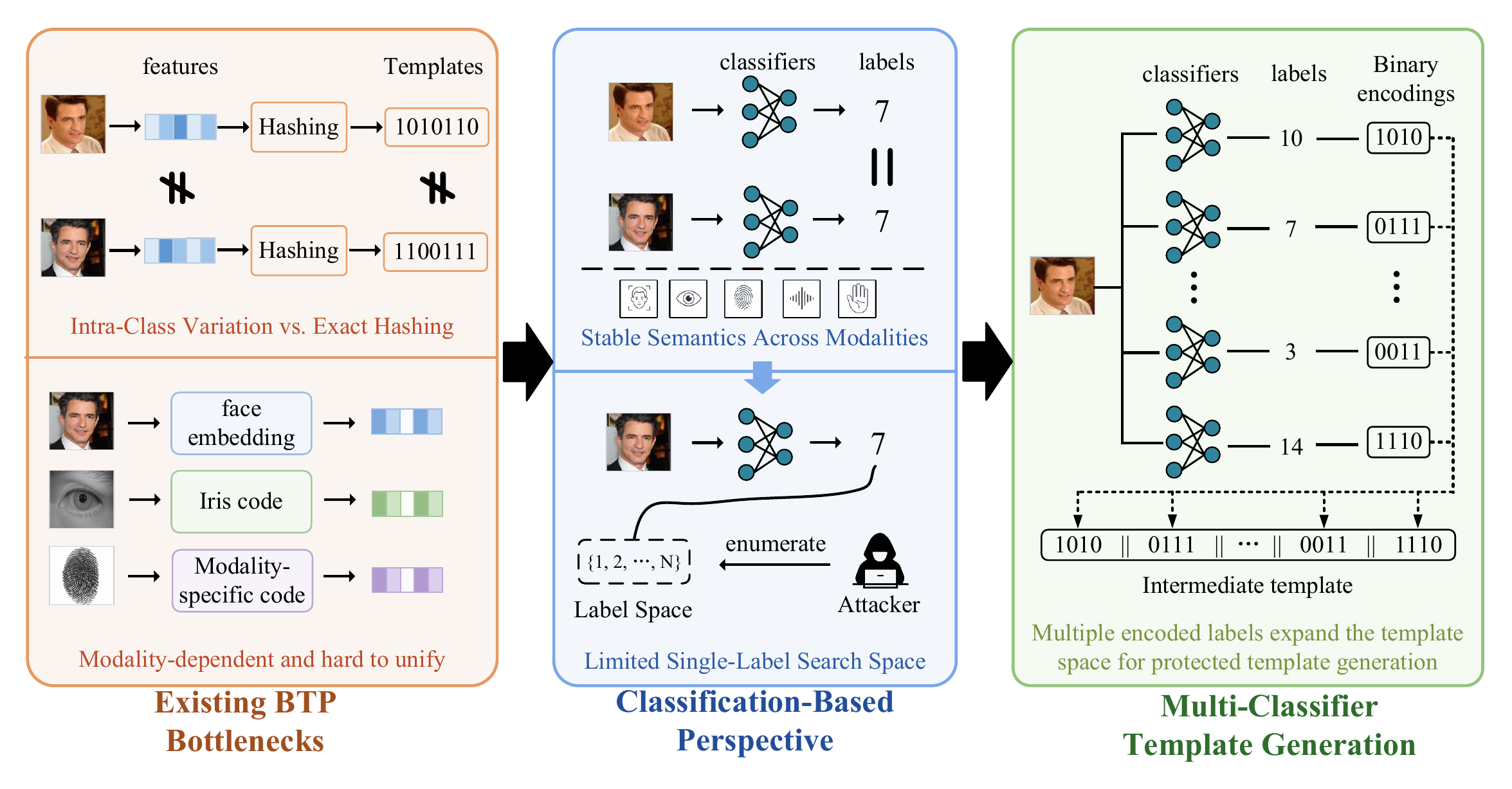}
    \caption{Motivation of MPLCE: addressing existing BTP bottlenecks through classification-based template generation.}
    \label{fig:motivation}
\end{figure*}

As illustrated in Fig.~\ref{fig:motivation}, we address this challenge by revisiting classification as a mechanism for generating stable intermediate templates. For a fixed enrolled user set, a classifier learns discriminative decision regions that map biometric samples to predicted labels. When these predictions remain stable under intra-class variations, samples from the same subject can be converted from variable biometric observations into consistent discrete outputs, which are more suitable as inputs to cryptographic hashing. Compared with templates constructed directly from continuous biometric features, classification outputs provide a more abstract label-level basis for template generation and reduce the dependence of protected template construction on modality-specific feature representations. However, directly using a single classification label as the intermediate template provides only a limited and easily enumerable candidate space. Therefore, the template generation mechanism should retain the consistency provided by identity classification while enlarging the candidate space of intermediate templates and removing the direct identity-to-label structure.

Motivated by this observation, we propose \textbf{Multiple Permuted-Label Classifier Encoding (MPLCE)}, a BTP framework for consistent intermediate template generation and protected template construction. MPLCE constructs multiple classifier-specific label permutations over a fixed enrolled identity set and trains one classifier under each permutation, such that the same identity is assigned different label indices across classifiers. For an input sample, the predicted permuted labels are encoded and concatenated in a fixed classifier order to form an intermediate template. The label permutations prevent the multi-classifier output from degenerating into repeated encodings of a single identity label, while the combination of multiple predictions enlarges the effective candidate space of the intermediate template and preserves classification-level consistency. The intermediate template is then randomized with an application-specific XOR string and processed by a cryptographic hash function to generate the stored protected template, enabling exact-match verification. Because the protection process operates on identity-level label outputs, modality-specific classifiers can be used for different biometric traits while retaining the same label-based template construction and protection procedure. MPLCE therefore provides a modality-agnostic template construction principle at the protection level.

The main contributions of this paper are summarized as follows:
\begin{itemize}
     \item We propose MPLCE, a BTP framework that uses identity classification to generate consistent intermediate templates and combines application-specific randomization with cryptographic hashing for exact-match verification without error correction codes or biometric-dependent helper data.

    \item We introduce a permuted-label multi-classifier mechanism that assigns different labels to the same identity across classifiers, preventing repeated identity-label encodings and enlarging the effective candidate space of intermediate templates while preserving classification-level consistency.

    \item We establish a modality-agnostic template construction principle at the protection level, allowing the same label-based template construction and protection procedure to be used with modality-specific classifiers.

    \item We conduct extensive experiments on four face and two iris datasets. The results demonstrate competitive protected verification performance and support the irreversibility, revocability, and unlinkability of MPLCE under the considered threat model.
\end{itemize}

\section{Related Work}
We review existing BTP methods in two groups: traditional BTP methods and NN-based BTP methods.

\subsection{Traditional BTP Methods}

CB methods protect biometric templates through revocable and noninvertible transformations, allowing new protected templates to be generated by changing the transformation parameters. Representative designs include random projection~\cite{ref6}, Bloom filters~\cite{ref9}, local ranking~\cite{ref7}, Index-of-Max (IoM) hashing~\cite{ref12}, and R$\cdot$HoG transformations~\cite{ref15}. These methods protect biometric representations through randomized projection, binary set encoding, local order relations, maximum-index encoding, or alignment-robust feature transformation, respectively. Recent studies have further combined partial Walsh transformations with SimHash for protected face template generation~\cite{ref_gao2024_simhash} and with MinHash for cancelable binary template generation~\cite{ref_song2025_minhash}. A recent benchmark evaluated representative CB schemes across multiple biometric modalities and showed that their recognition and privacy behavior depends on both the protection mechanism and the underlying modality~\cite{ref_cb_benchmark}. Despite their different designs, many CB methods retain matchable structures for protected-domain comparison. Security analyses of CB schemes have demonstrated that their structural information can be exploited under specific attack models~\cite{ref10,ref11}. Known-sample and similarity-based attacks further show that relations preserved in protected templates may facilitate template compromise or cross-record analysis~\cite{ref13,ref14,ref16}.

BCS combine biometric data with cryptographic constructions for key binding or key generation. Representative schemes include fuzzy vault~\cite{ref17,ref18}, fuzzy commitment~\cite{ref19,ref20}, and fuzzy extractor~\cite{ref21}. Fuzzy vault conceals a secret using genuine biometric feature points and chaff points, fuzzy commitment binds a biometric representation to an error-correcting codeword, and fuzzy extractor derives a reproducible cryptographic key from noisy biometric inputs with the aid of public helper data. These constructions typically rely on helper data and error correction or recovery mechanisms to tolerate intra-class variations and reproduce consistent keys or codewords. Recent studies have extended BCS to deep and multi-biometric representations. WiFaKey adapts deep face representations to error-correction requirements and employs neural decoding for key generation under unconstrained conditions~\cite{ref65}, while a recent deep multi-biometric fuzzy commitment scheme fuses fingerprint and iris embeddings using different fusion and error-correction strategies~\cite{ref_fohr2025_dmfc}. These studies also indicate that the effectiveness of BCS depends on the compatibility among biometric representations, error distributions, and recovery mechanisms. The security of BCS further depends on the specific construction and threat model. Fuzzy-vault schemes have been shown vulnerable to point-set recovery, cracking, and collusion attacks~\cite{ref23,ref24,ref25}. Fuzzy commitment schemes and biometric sketches may expose statistical or auxiliary information~\cite{ref26,ref27,ref51}, while the reuse of secure sketches or fuzzy extractors across records can introduce additional security risks~\cite{ref29}.

\subsection{BTP Methods Based on Neural Networks}

Existing NN-based BTP methods can be broadly divided according to the role of neural networks: some use NNs only for biometric feature extraction and rely on subsequent non-NN protection mechanisms, whereas others directly learn protected templates~\cite{ref4}. This subsection focuses on the latter category, which mainly includes methods that learn protected representations or stable codes and methods that learn mappings to predefined user templates.

The first category directly learns protected representations or stable codes from biometric samples~\cite{ref34,ref35,ref_biodeephash}. The method in~\cite{ref34} trains a network to approximate an IoM transformation and performs similarity matching on the resulting protected templates. The method in~\cite{ref35} combines a random convolutional neural network (CNN), random triplet loss, and user-specific keys to generate protected representations, while secure sketches are used to avoid directly storing the keys. BioDeepHash~\cite{ref_biodeephash} learns stable binary codes and further applies application-specific randomization and cryptographic hashing to generate the stored protected templates. Although their protection mechanisms differ, these methods all require the learned representation or code to remain sufficiently stable under intra-class variations. Residual differences are handled through protected-domain matching, secure sketches, or the direct generation of sufficiently consistent codes before hashing.

The second category assigns each enrolled user a predefined binary template or code and trains a network to reproduce the assigned target~\cite{ref32,ref33,ref62}. The method in~\cite{ref32} uses random maximum-entropy binary codes, whereas~\cite{ref33} adopts low-density parity-check (LDPC) codewords and exploits their error correction capability to improve robustness against intra-class variations. These methods produce user-specific discrete outputs suitable for subsequent cryptographic processing, but their template generation depends on learning a direct mapping from biometric samples to complete predefined codes.

Despite their different constructions, these methods generally require the network to learn or reproduce a complete protected representation or predefined code, with intra-class variations handled through protected-domain matching, secure sketches, error correction, or sufficiently stable output generation. In contrast, MPLCE uses identity classification to generate discrete label sequences, shifting the consistency requirement from reproducing a complete representation or code to maintaining stable identity predictions before subsequent template protection.

\begin{figure*}
    \centering
    \includegraphics[width=\textwidth]{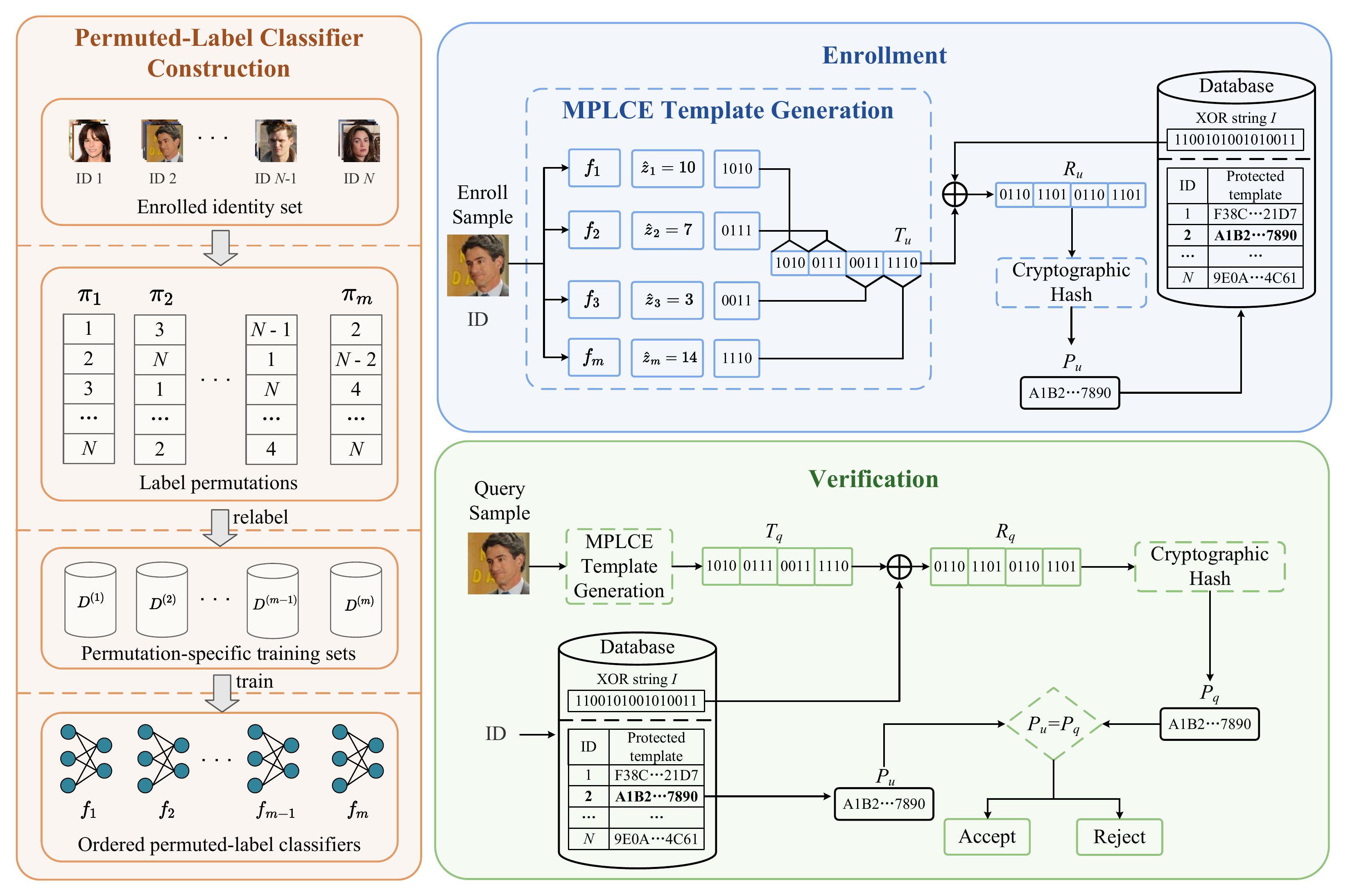}
    \caption{Overall framework of MPLCE for classifier construction, enrollment, and verification.}
    \label{fig:framework}
\end{figure*}

\section{MPLCE}

This section presents the MPLCE framework. We first provide an overview of permuted-label classifier construction, enrollment, and verification. We then formulate the consistency requirement of intermediate templates for hash-based verification and describe the generation of intermediate and protected templates together with the verification procedure.

\subsection{Overview of MPLCE}

Fig.~\ref{fig:framework} illustrates the overall workflow of MPLCE, which consists of three stages: permuted-label classifier construction, enrollment, and verification. In the classifier construction stage, MPLCE trains an ordered set of classifiers over classifier-specific permuted-label spaces. During enrollment, an enrollment sample is processed by the trained classifiers to generate an intermediate template, which is then randomized with an application-specific XOR string and processed by a cryptographic hash function to obtain the protected template. The database stores the registered identity index, the protected template, and the application-specific XOR string, whereas the intermediate template, the randomized template, and the original biometric sample are not stored.

During verification, a query sample is processed through the same template generation and protection procedure using the application-specific XOR string associated with the claimed identity. The resulting query protected template is compared with the stored protected template by exact matching. Because exact matching of cryptographic hash outputs requires identical hash inputs, genuine enrollment and query samples must generate consistent intermediate templates before hashing. Therefore, intra-class consistency and identity discrimination are central requirements for intermediate template generation in MPLCE, as formulated next.

\subsection{Formulation of Stable Intermediate Template Generation}

In MPLCE, protected templates generated after application-specific randomization and cryptographic hashing are verified by exact matching. This verification rule requires the pre-hash intermediate templates to be consistent for genuine samples and distinguishable across different users. Therefore, we formulate intermediate template generation as a problem of satisfying intra-class consistency and inter-class separability before protection.

Let \(X\) denote the biometric sample space, and let
\(Y=\{1,2,\ldots,N\}\) denote the enrolled user label space,
where \(N\) is the number of enrolled users.
For the enrolled user set, the enrollment data are denoted as
\(
D=\{(x_i,y_i)\}_{i=1}^{n},
\)
where \(x_i\in X\) is a biometric sample and \(y_i\in Y\) is its user label.
In MPLCE, the enrollment data are used to construct the permuted-label classifiers, and enrolled samples are used to generate the stored protected templates.
The objective is to construct an intermediate template generation function
\begin{equation}    
T: X\rightarrow\{0,1\}^{L},
\end{equation}
where \(L\) is the length of the generated intermediate template.

An ideal intermediate template generation function should satisfy two desired properties: intra-class consistency and inter-class separability. Given two biometric samples \(x_i\) and \(x_j\) with user labels \(y_i\) and \(y_j\), intra-class consistency requires genuine samples to produce identical intermediate templates:
\begin{equation}
T(x_i)=T(x_j), \quad \text{if } y_i=y_j.
\label{eq:intra_class_consistency}
\end{equation}
This property supports protected verification based on exact matching, because inconsistent intermediate templates from the same user lead to verification failure after protection. Inter-class separability requires samples from different users to produce different intermediate templates:
\begin{equation}
T(x_i)\neq T(x_j), \quad \text{if } y_i\neq y_j.
\label{eq:inter_class_separability}
\end{equation}
This property reduces false acceptance caused by intermediate-template collisions across users.

\subsection{Permuted-Label Multi-Classifier Template Generation}

MPLCE implements the intermediate template generation function \(T(\cdot)\) through a permuted-label multi-classifier mechanism. The key idea is to construct multiple classifier-specific label spaces over the same enrolled user label space, so that each classifier provides one permuted-label prediction for template generation. The following part formalizes label permutation, classifier construction, and intermediate template construction.

To construct multiple permuted-label classifiers, MPLCE first generates \(m\) label permutations over the enrolled user label space:
\begin{equation}
\pi_j: Y\rightarrow Y, \quad j=1,2,\ldots,m,
\label{eq:label_permutation}
\end{equation}
where each \(\pi_j(\cdot)\) is a bijective mapping over \(Y\). For the \(j\)-th permutation, each original user label \(y_i\) is replaced by its permuted label
\begin{equation}
z_i^{(j)}=\pi_j(y_i),
\label{eq:permuted_label}
\end{equation}
which defines the relabeled dataset for the \(j\)-th classifier:
\begin{equation}
D^{(j)}=\{(x_i,z_i^{(j)})\}_{i=1}^{n}.
\label{eq:permutation_training_set}
\end{equation}
Thus, different classifier-specific label spaces are constructed by changing only the user label mappings, while the biometric samples themselves remain unchanged.

For each relabeled dataset \(D^{(j)}\), MPLCE trains one permuted-label classifier \(f_j(\cdot;\theta_j)\). The classifiers share the same classification architecture but are optimized under different label mappings. The training objective of the \(j\)-th classifier is formulated as
\begin{equation}
\theta_j^{*}
=
\arg\min_{\theta_j}
\sum_{i=1}^{n}
\mathcal{L}_{\mathrm{cls}}
\left(
f_j(x_i;\theta_j), z_i^{(j)}
\right),
\label{eq:classifier_training}
\end{equation}
where \(\mathcal{L}_{\mathrm{cls}}(\cdot)\) denotes the classification loss used for training the permuted-label classifier. After training all \(m\) classifiers, MPLCE obtains an ordered classifier sequence \(\mathcal{F}=(f_1(\cdot;\theta_1^{*}), f_2(\cdot;\theta_2^{*}), \ldots, f_m(\cdot;\theta_m^{*}))\). The order of \(\mathcal{F}\) is fixed and used consistently during enrollment and verification for intermediate template construction.

\begin{algorithm}[t]
\caption{MPLCE Classifier Construction and Intermediate Template Generation}
\label{algo:multi_classifier_generation}
\KwIn{
training set \(D=\{(x_i,y_i)\}_{i=1}^{n}\),
enrolled identity label space \(Y\),
number of classifiers \(m\),
classifier architecture \(f(\cdot;\theta)\),
label encoding function \(\mathrm{Enc}(\cdot)\),
input sample \(x\)
}
\KwOut{
ordered classifier sequence \(\mathcal{F}=(f_1,\ldots,f_m)\),
intermediate template \(T(x)\)
}

\BlankLine
\textbf{Permuted-label classifier construction:}\\
\(\mathcal{F}\leftarrow [\,]\)\;

\For{\(j\leftarrow 1\) \KwTo \(m\)}{
    Generate a bijective label permutation \(\pi_j:Y\rightarrow Y\)\;
    \(D^{(j)}\leftarrow \emptyset\)\;

    \For{\(i\leftarrow 1\) \KwTo \(n\)}{
        \(z_i^{(j)}\leftarrow \pi_j(y_i)\)\;
        \(D^{(j)}\leftarrow D^{(j)}\cup\{(x_i,z_i^{(j)})\}\)\;
    }

    Initialize \(f_j(\cdot;\theta_j)\) with architecture \(f(\cdot;\theta)\)\;
    Train \(f_j(\cdot;\theta_j)\) on \(D^{(j)}\) and obtain \(\theta_j^{*}\)\;
    \(\mathcal{F}[j]\leftarrow f_j(\cdot;\theta_j^{*})\)\;
}

\BlankLine
\textbf{Intermediate template generation for sample \(x\):}\\
\For{\(j\leftarrow 1\) \KwTo \(m\)}{
    \(\hat{z}_j(x)\leftarrow
    \arg\max_{k\in Y}
    \left[f_j(x;\theta_j^{*})\right]_k\)\;
    \(b_j(x)\leftarrow \mathrm{Enc}\left(\hat{z}_j(x)\right)\)\;
}

\(T(x)\leftarrow b_1(x)\Vert b_2(x)\Vert\cdots\Vert b_m(x)\)\;

\Return \(\mathcal{F}\), \(T(x)\)\;
\end{algorithm}

Given an input sample \(x\), each trained classifier \(f_j(\cdot;\theta_j^{*})\) predicts a label in its own permuted-label space. The predicted label of the \(j\)-th classifier is defined as
\begin{equation}
\hat{z}_j(x)
=
\arg\max_{k\in Y}
\left[f_j(x;\theta_j^{*})\right]_k ,
\label{eq:predicted_label}
\end{equation}
where \(\left[f_j(x;\theta_j^{*})\right]_k\) denotes the classification score for label \(k\) in the \(j\)-th permuted-label space.

The predicted label is then converted into a fixed-length binary segment. Let
\begin{equation}
\ell=\lceil \log_2 N\rceil .
\label{eq:label_bit_length}
\end{equation}
For a label \(k\in Y\), the label encoding function is defined as
\begin{equation}
\mathrm{Enc}(k)=(e_{\ell-1}(k),e_{\ell-2}(k),\ldots,e_0(k)),
\label{eq:label_encoding_function}
\end{equation}
where
\begin{equation}
e_r(k)=
\left\lfloor \frac{k-1}{2^r} \right\rfloor \bmod 2,
\quad r=0,1,\ldots,\ell-1.
\label{eq:label_encoding_bit}
\end{equation}
Thus, \(\mathrm{Enc}(k)\) is the \(\ell\)-bit binary representation of \(k-1\), ordered from the most significant bit to the least significant bit. The binary segment generated by the \(j\)-th classifier is
\begin{equation}
b_j(x)
=
\mathrm{Enc}\left(\hat{z}_j(x)\right),
\quad
b_j(x)\in\{0,1\}^{\ell}.
\label{eq:label_encoding}
\end{equation}

The binary segments generated by all classifiers are concatenated in the fixed classifier order to form the intermediate template:
\begin{equation}
T(x)
=
b_1(x)\Vert b_2(x)\Vert\cdots\Vert b_m(x),
\label{eq:intermediate_template}
\end{equation}
where \(\Vert\) denotes concatenation. The resulting intermediate template has a binary length of \(L=m\ell\). However, not every \(L\)-bit string corresponds to a valid intermediate template in MPLCE, because each segment must encode one of the \(N\) valid labels. Therefore, the effective candidate space of the intermediate template contains \(N^m\) candidates rather than all \(2^L\) binary strings. The same classifier order is used during enrollment and verification, ensuring that identical predicted label sequences produce identical intermediate templates.

Algorithm~\ref{algo:multi_classifier_generation} summarizes the construction of the permuted-label classifiers and the generation of intermediate templates.

\subsection{Protected Template Generation and Exact-Match Verification}

Given the intermediate template \(T(x)\), MPLCE applies an application-specific XOR string \(I\) before cryptographic hashing:
\begin{equation}
R(x)=T(x)\oplus I,
\label{eq:randomized_template_general}
\end{equation}
where \(I\in\{0,1\}^{L}\) and \(\oplus\) denotes bitwise XOR. The XOR string \(I\) is generated independently of biometric data and is not treated as secret information. The randomized template is then processed by a cryptographic hash function \(h(\cdot)\) to generate the protected template:
\begin{equation}
P(x)=h(R(x))=h(T(x)\oplus I).
\label{eq:protected_template_general}
\end{equation}
Both \(T(x)\) and \(R(x)\) are temporary values and are discarded after \(P(x)\) is generated.

For an enrolled user \(u\) with enrollment sample \(x_u\), the stored protected template is denoted by \(P_u=P(x_u)\). The protected-template database stores the registered user index, \(P_u\), and the application-specific XOR string \(I\), but does not retain the intermediate template \(T(x_u)\), the randomized template \(R(x_u)\), or the enrollment sample \(x_u\). Because \(I\) is generated independently of biometric data, its storage does not constitute biometric-dependent helper data.

During verification, given a query sample \(x_q\) and a claimed user \(u\), the system retrieves \(P_u\) and \(I\) from the corresponding protected record. The query sample is processed using the same template generation and protection procedure:
\begin{equation}
P_q=h(T(x_q)\oplus I).
\label{eq:query_protected_template}
\end{equation}
The verification decision is made by exact matching:
\begin{equation}
\mathrm{Verify}(x_q,u)
=
\mathbf{1}\left[P_q=P_u\right],
\label{eq:exact_match_verification}
\end{equation}
where \(\mathbf{1}[\cdot]\) denotes the indicator function. The query is accepted only when the query and enrolled protected templates are identical.

\section{Security Analysis}

\subsection{Threat Model}
To analyze the security of MPLCE, we adopt the full-disclosure model specified in ISO/IEC 30136:2018, which assumes the strongest adversarial knowledge considered in BTP evaluation~\cite{ref_iso30136}. In this model, the protection mechanism is assumed to be public, and the adversary may obtain compromised protected records. The adversary's goals and knowledge are defined as follows.

\begin{itemize}

\item \textbf{Adversary's goals}: The adversary aims to compromise the security or privacy of an enrolled user by exploiting compromised protected records. The main concerns are whether the compromised records can be used to generate accepted impostor attempts, recover biometric-related information, or determine whether protected templates from different applications belong to the same user.

\item \textbf{Adversary's knowledge}: The adversary is assumed to know the public MPLCE specification, including the template protection workflow, the construction principle, architectures, and trained parameters of the permuted-label classifiers, the label encoding rule, the XOR randomization operation, the cryptographic hash function, and the exact-match verification rule.

\end{itemize}

In addition to this public knowledge, the adversary may obtain the records stored in the protected template database, including the protected templates \(P_u\) and their corresponding application-specific XOR strings \(I\). Therefore, \(I\) is not assumed to be secret. The compromised database records do not contain the enrolled intermediate templates \(T_u\), the randomized hash inputs \(T_u\oplus I\), the original enrollment samples, continuous biometric feature representations, or the concrete permutation tables used to instantiate the classifier-specific label assignments. The concrete permutation tables are not included in the protected-template records and, because they are generated independently of biometric information, do not constitute biometric-dependent helper data.

\subsection{Irreversibility Analysis}

Irreversibility requires that an adversary cannot feasibly recover the enrolled intermediate template or the original biometric information from a compromised protected template. In MPLCE, the protected template of user \(u\) is generated by applying cryptographic hashing to the randomized intermediate template. Under the threat model defined above, the adversary may obtain both the protected template \(P_u\) and the corresponding application-specific XOR string \(I\). Even in this case, the enrolled intermediate template \(T_u\) is not directly exposed, because the hash input \(T_u \oplus I\) is only used during protected template generation and is not retained.

Given that neither \(T_u\) nor \(T_u \oplus I\) is available from the compromised record, recovering \(T_u\) from \(P_u\), even when \(I\) is known, would require recovering the randomized hash input from the cryptographic hash output. Under the standard preimage-resistance assumption of the hash function, this recovery is computationally infeasible. If the adversary does not directly invert the hash function, an alternative strategy is to enumerate candidate intermediate templates and test whether they produce the compromised protected template. The difficulty of this enumeration-based recovery path depends on the candidate space induced by the multiple permuted-label classifiers, and is further quantified in Section~\ref{sec:representative_attacks}.

\subsection{Revocability Analysis}

Revocability requires that a compromised protected record can be invalidated and replaced without changing the user's biometric trait. In MPLCE, revocability is achieved by renewing the application-specific XOR string. Let \(T_u\) denote the enrolled intermediate template of an arbitrary user \(u\). When the application uses the XOR string \(I\), the protected template of user \(u\) is
\(
P_u=h(T_u\oplus I).
\)
After the application-specific XOR string is renewed from \(I\) to \(I'\), where \(I'\neq I\), the protected template of the same user is reissued as
\(
P'_u=h(T_u\oplus I').
\)
Since the hash input changes from \(T_u\oplus I\) to \(T_u\oplus I'\), the resulting protected template differs from the compromised one except with negligible hash-collision probability. This process does not require changing the user's biometric trait or retraining the MPLCE classifiers; it only requires generating a new application-specific XOR string and reissuing the protected template through the enrollment pipeline.

\subsection{Unlinkability Analysis}

Unlinkability requires that an adversary cannot reliably determine whether protected templates generated in different applications belong to the same enrolled user. In MPLCE, this property is supported by applying application-specific XOR randomization before cryptographic hashing.

Consider user \(u\) enrolled in two different applications \(s\) and \(t\). Let \(T_u\) denote the intermediate template generated for this user. The corresponding protected templates are
\begin{equation}
P_u^{(s)} = h(T_u \oplus I_s), \quad
P_u^{(t)} = h(T_u \oplus I_t),
\end{equation}
where \(I_s\) and \(I_t\) are independently generated application-specific XOR strings.

Under the full-disclosure model, the application-specific XOR strings may be known to the adversary. However, knowing \(I_s\) and \(I_t\) does not by itself allow the adversary to derive the underlying intermediate template or transform \(P_u^{(s)}\) and \(P_u^{(t)}\) into a directly comparable representation, because the randomized hash inputs are hidden by the cryptographic hash function. Recovering these hash inputs from the protected templates would require inverting the cryptographic hash function, which is computationally infeasible under the preimage-resistance assumption of \(h(\cdot)\). Therefore, protected templates generated in different applications do not expose a directly comparable intermediate template representation. This property is further evaluated in Section~\ref{sec:representative_attacks}, where mated and non-mated cross-application protected template pairs are compared using standard unlinkability metrics.

\subsection{Discussion of Theoretical Security Boundaries}

MPLCE comprises two functionally distinct stages. The classifier stage generates intermediate templates from biometric samples, whereas the protection stage applies XOR randomization and cryptographic hashing to generate the protected records stored in the database. The analysis in this section addresses the security of these stored records after compromise, specifically their irreversibility, revocability, unlinkability, and resistance to offline enumeration.

This boundary is defined by the security object under analysis rather than by restricting the adversary's knowledge. As specified in the threat model, the adversary is assumed to know the architectures and trained parameters of the permuted-label classifiers, together with the complete template protection procedure. Nevertheless, model-level vulnerabilities that affect the correctness or confidentiality of intermediate template generation constitute a separate attack surface. Such vulnerabilities may affect the security of the overall biometric system, but they are not direct consequences of compromising the stored protected records. Accordingly, the security conclusions in this section apply to the protection of the stored records and should not be interpreted as comprehensive security guarantees for every component of the biometric recognition system.

\subsection{Security Evaluation Against Representative Attacks}
\label{sec:representative_attacks}

To complement the preceding analysis, we evaluate MPLCE against three representative attack categories: brute-force attacks, similarity-based attacks, and attacks via record multiplicity (ARM).

\subsubsection{Brute-Force Attack}

Brute-force resistance in BTP is commonly characterized by the effort required to search for a preimage or an equivalent template that satisfies the system's matching rule~\cite{ref42}. Under the defined threat model, we consider an offline attack against MPLCE in which the adversary enumerates candidate intermediate templates and tests whether the resulting protected templates match any record in the compromised database. Given the stored protected templates and the application-specific XOR string \(I\), the adversary computes
\(\widetilde{P}=h(\widetilde{T}\oplus I)\)
for each candidate intermediate template \(\widetilde{T}\) and compares it with the stored protected templates.

For an enrolled identity set of size \(N\), each of the \(m\) classifiers contributes one label with \(N\) possible values. Since the fixed-length label encoding is injective, every valid label sequence corresponds to one candidate intermediate template. Therefore, the offline candidate space under the defined threat model is
\begin{equation}
|\Omega_T|=N^m,
\end{equation}
where \(\Omega_T\) denotes the candidate intermediate templates considered in offline enumeration. This space contains only valid encoded label sequences rather than all \(2^L\) binary strings.

We further perform randomized brute-force search under each \((N,m)\) setting to evaluate random guessing under a finite budget. In each trial, a candidate is sampled uniformly from \(\Omega_T\), converted into an intermediate template, randomized using \(I\), and hashed for comparison with all protected templates in the compromised database. Since each of the \(N\) enrolled identities has one stored protected template, the per-guess success probability of this untargeted attack is at most \(N/|\Omega_T|\). For \(G_{\mathrm{BF}}\) independent guesses with replacement, the success probability is bounded by
\begin{equation}
P_{\mathrm{BF}}
\leq
1-\left(1-\frac{N}{|\Omega_T|}\right)^{G_{\mathrm{BF}}}.
\end{equation}

Table~\ref{tab:bf_results} reports the candidate-space size, the finite-budget success probability upper bound, and the observed number of successful matches under \(G_{\mathrm{BF}}=10\textsuperscript{9}\). Succ. denotes the number of successful matches observed during randomized brute-force search. No successful match is observed in any evaluated setting. This result is consistent with the theoretical upper bounds and indicates that offline brute-force search is practically infeasible under the considered guessing budget.

\begin{table}[t]
\centering
\caption{Candidate-space analysis and random-search results for untargeted offline enumeration with \(G_{\mathrm{BF}}=10^9\).}
\label{tab:bf_results}
\small
\renewcommand{\arraystretch}{1.05}
\begin{tabular*}{\columnwidth}
{@{\hspace{4pt}}c@{\extracolsep{\fill}}cccc@{\hspace{4pt}}}
\toprule
\multirow{2}{*}{\(N\)}
& \multirow{2}{*}{\(m\)}
& \multicolumn{2}{c}{\textbf{Theoretical analysis}}
& \multicolumn{1}{c}{\textbf{Random search}} \\
\cmidrule(lr){3-4}
\cmidrule(lr){5-5}
& & \(|\Omega_T|\) & \(P_{\mathrm{BF}}\) & \textbf{Succ.} \\
\midrule

\multirow{3}{*}{100}
& 12
& 1.00\(\times\)10\textsuperscript{24}
& \(\leq\) 1.00\(\times\)10\textsuperscript{-13}
& 0 \\

& 16
& 1.00\(\times\)10\textsuperscript{32}
& \(\leq\) 1.00\(\times\)10\textsuperscript{-21}
& 0 \\

& 20
& 1.00\(\times\)10\textsuperscript{40}
& \(\leq\) 1.00\(\times\)10\textsuperscript{-29}
& 0 \\
\midrule

\multirow{3}{*}{200}
& 12
& 4.10\(\times\)10\textsuperscript{27}
& \(\leq\) 4.88\(\times\)10\textsuperscript{-17}
& 0 \\

& 16
& 6.55\(\times\)10\textsuperscript{36}
& \(\leq\) 3.05\(\times\)10\textsuperscript{-26}
& 0 \\

& 20
& 1.05\(\times\)10\textsuperscript{46}
& \(\leq\) 1.91\(\times\)10\textsuperscript{-35}
& 0 \\
\midrule

\multirow{3}{*}{300}
& 12
& 5.31\(\times\)10\textsuperscript{29}
& \(\leq\) 5.65\(\times\)10\textsuperscript{-19}
& 0 \\

& 16
& 4.30\(\times\)10\textsuperscript{39}
& \(\leq\) 6.97\(\times\)10\textsuperscript{-29}
& 0 \\

& 20
& 3.49\(\times\)10\textsuperscript{49}
& \(\leq\) 8.60\(\times\)10\textsuperscript{-39}
& 0 \\
\bottomrule
\end{tabular*}
\end{table}

\subsubsection{Similarity-Based Attack}

A similarity-based attack exploits meaningful protected-domain distances to guide the search for a biometric preimage or pseudo-template~\cite{ref14}. Since MPLCE performs exact matching on cryptographic hash outputs, it does not provide such a similarity score. We therefore evaluate the key prerequisite of this attack category by examining whether distances between stored protected templates remain correlated with distances in the unprotected biometric feature space.

We use VGGFace2 with IR101 and CASIA-Iris-Lamp with ResNet18 as representative face and iris settings, respectively, with \(N=200\). For a non-mated pair \((x_i,x_j)\), the unprotected feature distance is
\begin{equation}
d_{\mathrm{feat}}(x_i,x_j)=1-\operatorname{cos}(e_i,e_j),
\end{equation}
where \(e_i\) and \(e_j\) are the corresponding biometric features. The protected templates are generated using the same application-specific XOR string, and their normalized Hamming distance is
\begin{equation}
d_{\mathrm{prot}}(P_i,P_j)=\frac{d_H(P_i,P_j)}{\lambda},
\end{equation}
where \(\lambda=512\) for SHA3-512. Although Hamming distance is not used for MPLCE verification, it is employed here to diagnose potential distance leakage. For each setting, we sample \(10\textsuperscript{5}\) non-mated pairs and report the Pearson and Spearman correlations between \(d_{\mathrm{feat}}\) and \(d_{\mathrm{prot}}\), together with the mean normalized Hamming distance. As shown in Table~\ref{tab:similarity_leakage}, all correlations are close to zero, with a maximum absolute value of 0.0122, while the mean normalized Hamming distance remains approximately 0.5. These results indicate that, under the evaluated settings and metrics, the stored protected templates do not provide an observable distance signal for guiding similarity-based attacks.

\begin{table}[t]
\centering
\caption{Distance correlation analysis on representative face and iris settings under \(N=200\).}
\label{tab:similarity_leakage}
\small
\renewcommand{\arraystretch}{1.05}
\begin{tabular*}{\columnwidth}
{@{\hspace{4pt}}c@{\extracolsep{\fill}}cccc@{\hspace{4pt}}}
\toprule
\textbf{Dataset}
& \(m\)
& \textbf{Pearson} \(r\)
& \textbf{Spearman} \(\rho\)
& \(\overline{d}_{\mathrm{prot}}\) \\
\midrule

\multirow{3}{*}{VGGFace2}
& 12 & 0.0030 & 0.0038 & 0.5000 \\
& 16 & 0.0107 & 0.0122 & 0.5001 \\
& 20 & 0.0042 & 0.0007 & 0.4998 \\
\midrule

\multirow{3}{*}{Lamp}
& 12 & -0.0005 & -0.0053 & 0.4999 \\
& 16 & 0.0023 & -0.0011 & 0.4997 \\
& 20 & 0.0050 & 0.0058 & 0.5000 \\
\bottomrule
\end{tabular*}
\end{table}

\begin{figure*}[t]
\centering
\subfloat[VGGFace2, $m=12$]{
    \includegraphics[width=0.3\textwidth]{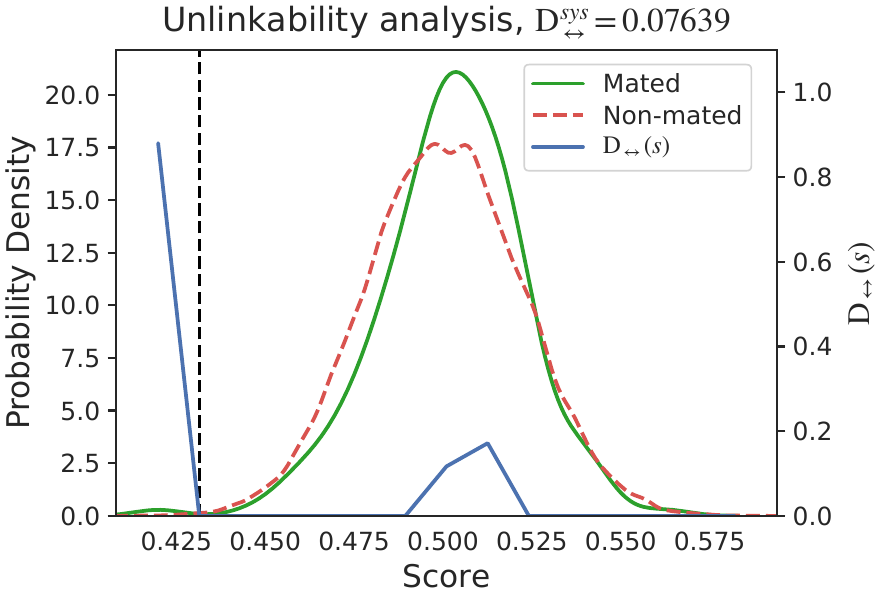}
}
\hfill
\subfloat[VGGFace2, $m=16$]{
    \includegraphics[width=0.3\textwidth]{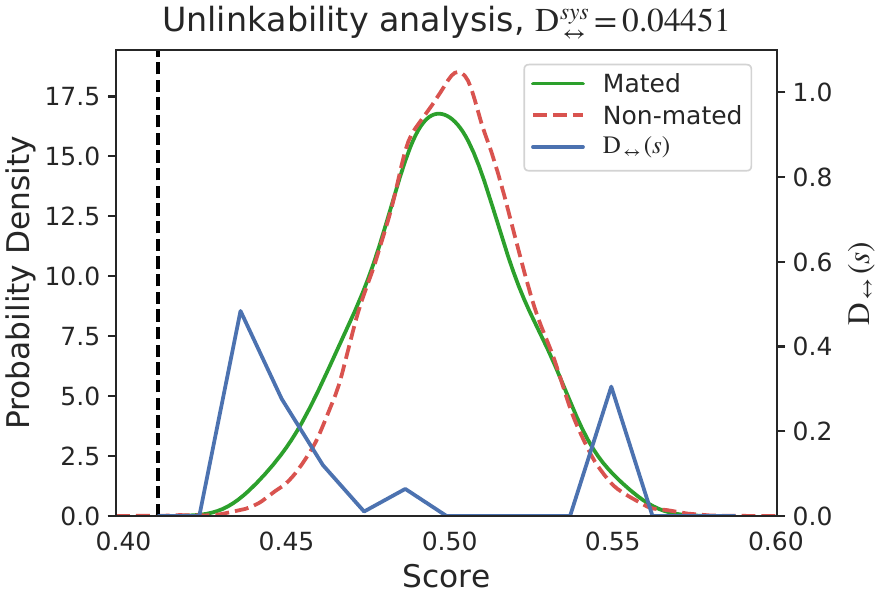}
}
\hfill
\subfloat[VGGFace2, $m=20$]{
    \includegraphics[width=0.3\textwidth]{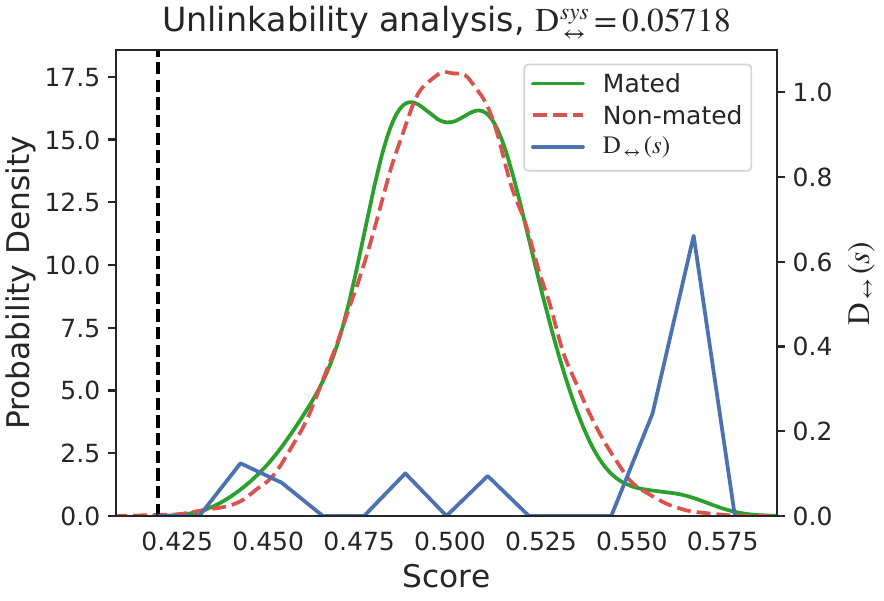}
}

\vspace{0.5em}

\subfloat[Lamp, $m=12$]{
    \includegraphics[width=0.3\textwidth]{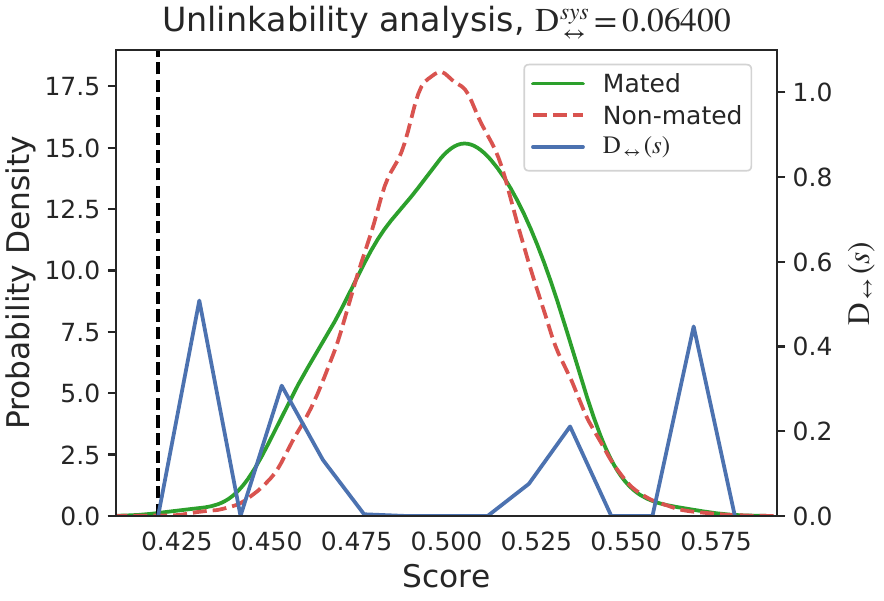}
}
\hfill
\subfloat[Lamp, $m=16$]{
    \includegraphics[width=0.3\textwidth]{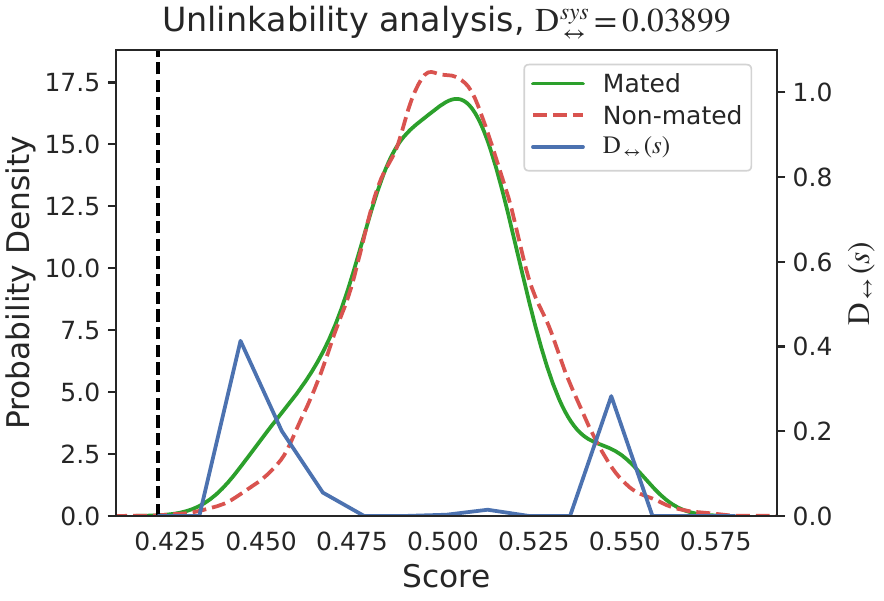}
}
\hfill
\subfloat[Lamp, $m=20$]{
    \includegraphics[width=0.3\textwidth]{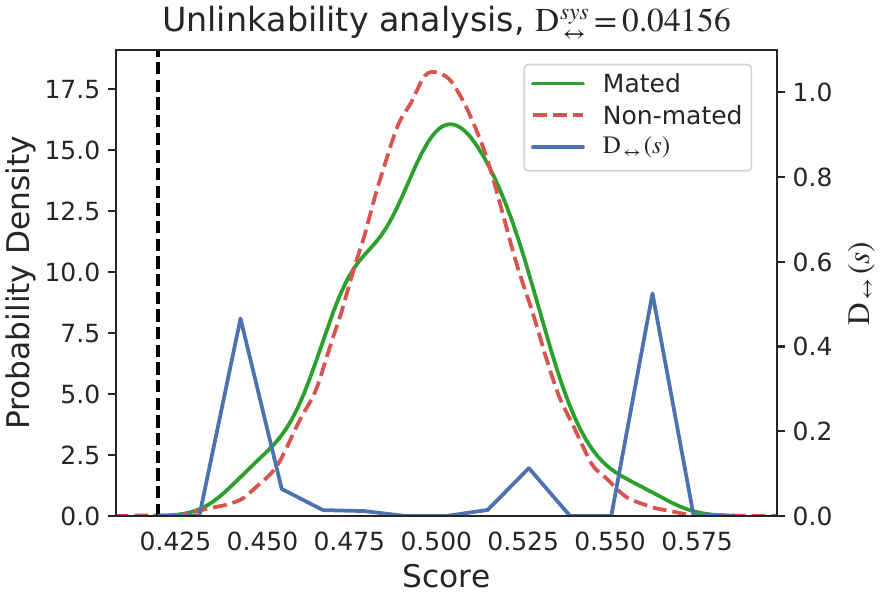}
}

\caption{Unlinkability evaluation of MPLCE on VGGFace2 and CASIA-Iris-Lamp under different numbers of classifiers.}
\label{fig:unlinkability_2x3}
\end{figure*}

\subsubsection{Attacks via Record Multiplicity}
\label{sec:arm}

We evaluate ARM as a cross-application record linkage attack using the unlinkability framework in~\cite{ref41}. Mated cross-application pairs are constructed from protected templates generated for the same user in different applications, whereas non-mated pairs are constructed from templates generated for different users in different applications. The framework compares the score distributions of these two types of pairs and computes the local unlinkability measure \(D_{\leftrightarrow}(s)\) and the global system-level measure \(D_{\leftrightarrow}^{\mathrm{sys}}\). Both measures range from 0 to 1, with values closer to 0 indicating stronger unlinkability. Following this framework, we use a similarity score derived from the normalized Hamming distance between protected templates, where larger values indicate a higher likelihood of a mated pair.

As shown in Fig.~\ref{fig:unlinkability_2x3}, the mated and non-mated score distributions substantially overlap across all evaluated settings. The values of \(D_{\leftrightarrow}^{\mathrm{sys}}\) range from 0.03899 to 0.07639, indicating limited distinguishability between the two types of cross-application pairs. Therefore, under the evaluated settings, the protected records do not provide a reliable signal for ARM-based cross-application linkage.

\section{Experiments}




\subsection{Experimental Setup}

\subsubsection{Datasets and Data Splits}
We evaluate MPLCE on both face and iris modalities. For face, VGGFace2~\cite{ref_vggface2} and FaceScrub~\cite{ref_facescrub} are used for controlled evaluations across different values of \(N\) and \(m\) and multiple backbone models. YouTube Faces (YTF)~\cite{ref_ytf} and Extended Yale B (Yale-B)~\cite{ref_yaleb} are used for comparisons with existing BTP methods under aligned benchmark protocols. For iris, we use CASIA-Iris-Lamp (Lamp) and CASIA-Iris-Thousand (Thousand). The left and right irises of each subject are treated as separate identity classes.

For each experimental setting, \(N\) enrolled identity classes are selected according to the protocol specified in the corresponding subsection. For each class, \(K\) samples are used for both permuted-label classifier construction and template enrollment. After classifier training, these \(K\) samples are processed by the ordered classifiers, and the most frequently occurring intermediate template is used to generate the single protected template stored for that class. Query samples are excluded from classifier construction and template enrollment and are used only for verification. The specific values of \(N\), \(m\), and \(K\), together with the backbone models, data splits, and comparison protocols, are provided in the corresponding experimental subsections.

\subsubsection{Implementation Details}

SHA3-512 is used throughout all experiments. For face, we use IR50 and IR101 AdaFace models pretrained on MS1MV2~\cite{ref_adaface}, and an Inception-ResNet-v1 FaceNet model pretrained on CASIA-WebFace~\cite{ref_facenet}. The IR50 and IR101 models take \(112\times112\) images as input, with the last four backbone blocks and classifier head fine-tuned using learning rates of \(10^{-4}\) and \(10^{-3}\), respectively. FaceNet takes \(160\times160\) images as input and is fully fine-tuned using learning rates of \(10^{-5}\) for the backbone and \(10^{-3}\) for the classifier head. All face classifiers use Adam with a batch size of 64 and zero weight decay. IR50 and IR101 are trained for 20 epochs, and FaceNet for 30 epochs.

For iris, we use ResNet18, ResNet34~\cite{ref_resnet}, and DeepIrisNet~\cite{ref_deepirisnet}, all trained from scratch on \(64\times64\) images. The classifiers use Adam with an initial learning rate of \(3\times10^{-4}\), a batch size of 32, and zero weight decay. Cross-entropy and center loss~\cite{ref_centerloss} are jointly used, with both the center loss weight and center parameter learning rate set to \(10^{-3}\). ResNet18 and ResNet34 are trained for 30 epochs, and DeepIrisNet for 60 epochs, with the learning rate reduced by a factor of 0.1 halfway through training.

Within each dataset--backbone setting, the permuted-label classifiers share the same architecture and training configuration but are independently trained under different label permutations. When multiple values of \(m\) are compared within the same experimental setting, we train a single ordered classifier set using the largest tested value of \(m\) and use its first \(m\) classifiers for each configuration. This shared-prefix design reduces variations caused by independently trained classifier sets, so the differences across configurations mainly reflect the number of classifier outputs used for intermediate template construction. The enrolled intermediate template is regenerated for each \(m\). In each run, the permutations are generated once and fixed throughout training, enrollment, and verification. For each application, an application-specific XOR string \(I\in\{0,1\}^{L}\) is uniformly sampled and shared across its protected records. Unless otherwise specified, results are averaged over three independent runs. Models in different experimental subsections are trained independently, so identical \(N\), \(m\), and \(K\) settings may produce small variations across tables.

\subsubsection{Evaluation Protocol and Metrics}

During evaluation, each query sample is converted into a query protected template using the same MPLCE template generation and protection procedure. The query protected template is compared with the stored protected templates in the enrolled database by exact matching. The comparison with the stored protected template of the same identity is counted as a genuine attempt, while comparisons with the stored protected templates of other identities are counted as impostor attempts. A comparison is accepted only when the two protected templates are identical, and rejected otherwise.

We report verification performance using Genuine Acceptance Rate (GAR) and False Acceptance Rate (FAR). Since MPLCE performs exact-match verification in the protected domain, each comparison produces a binary accept/reject decision without a tunable threshold. Let \(TP\) denote the number of accepted genuine attempts, \(FN\) denote the number of rejected genuine attempts, \(FP\) denote the number of accepted impostor attempts, and \(TN\) denote the number of rejected impostor attempts. GAR and FAR are defined as
\begin{equation}
    \mathrm{GAR}=\frac{TP}{TP+FN}, \quad
    \mathrm{FAR}=\frac{FP}{FP+TN}.
\end{equation}
Both GAR and FAR are reported as percentages throughout this paper.

\begin{table*}[t]
\centering
\caption{Protected verification performance of MPLCE on face and iris datasets under different \(N\) and \(m\). Each entry reports GAR@FAR (\%).}
\label{tab:main_performance}
\scriptsize
\setlength{\tabcolsep}{2.6pt}
\renewcommand{\arraystretch}{1.15}
\begin{tabular}{@{}llccccccccc@{}}
\toprule
\multirow{2}{*}{\textbf{Dataset}}
& \multirow{2}{*}{\textbf{Backbone}}
& \multicolumn{3}{c}{{  \(N=100\)}}
& \multicolumn{3}{c}{{  \(N=200\)}}
& \multicolumn{3}{c}{{  \(N=300\)}} \\
\cmidrule(lr){3-5}
\cmidrule(lr){6-8}
\cmidrule(lr){9-11}
&
& {  \(m=12\)}
& {  \(m=16\)}
& {  \(m=20\)}
& {  \(m=12\)}
& {  \(m=16\)}
& {  \(m=20\)}
& {  \(m=12\)}
& {  \(m=16\)}
& {  \(m=20\)} \\
\midrule

\multirow{3}{*}{VGGFace2}
& IR50
& 94.35@1.01e-3
& 93.85@5.05e-4
& 93.40@5.05e-4
& 92.17@7.54e-4
& 91.57@6.28e-4
& 90.95@3.77e-4
& 91.27@6.13e-4
& 90.47@5.02e-4
& 89.82@3.34e-4 \\

& IR101
& 94.40@1.01e-3
& 93.85@1.01e-3
& 93.40@1.01e-3
& 92.55@6.28e-4
& 91.92@5.03e-4
& 91.40@5.03e-4
& 91.72@5.57e-4
& 91.12@3.34e-4
& 90.53@3.34e-4 \\

& FaceNet
& 93.40@7.58e-3
& 92.95@5.05e-3
& 92.50@5.05e-3
& 90.45@7.54e-3
& 90.22@6.66e-3
& 89.80@5.90e-3
& 89.43@5.63e-3
& 88.95@4.96e-3
& 88.57@4.74e-3 \\
\midrule

\multirow{3}{*}{FaceScrub}
& IR50
& 89.20@2.53e-3
& 87.95@1.52e-3
& 86.85@1.52e-3
& 85.70@1.01e-3
& 84.52@8.79e-4
& 83.47@6.28e-4
& 84.95@3.90e-4
& 83.63@2.79e-4
& 82.47@2.23e-4 \\

& IR101
& 89.95@1.01e-3
& 88.95@1.01e-3
& 87.95@1.01e-3
& 86.85@5.03e-4
& 85.78@2.51e-4
& 84.82@2.51e-4
& 85.63@3.90e-4
& 84.40@3.34e-4
& 83.30@2.23e-4 \\

& FaceNet
& 90.25@1.52e-2
& 89.60@1.46e-2
& 89.05@1.46e-2
& 87.20@7.16e-3
& 86.48@5.78e-3
& 85.90@5.03e-3
& 86.15@4.29e-3
& 85.53@3.79e-3
& 84.98@3.29e-3 \\
\midrule

\multirow{3}{*}{Thousand}
& ResNet18
& 94.00@5.05e-3
& 93.00@5.05e-3
& 92.00@5.05e-3
& 92.75@2.51e-3
& 92.25@2.51e-3
& 91.75@2.51e-3
& 91.17@1.67e-3
& 90.00@1.67e-3
& 89.17@1.67e-3 \\

& ResNet34
& 93.00@0.00
& 92.50@0.00
& 91.00@0.00
& 89.50@1.26e-3
& 88.25@1.26e-3
& 87.25@1.26e-3
& 88.67@2.23e-3
& 87.50@2.23e-3
& 86.50@1.67e-3 \\

& DeepIrisNet
& 94.00@0.00
& 93.00@0.00
& 92.00@0.00
& 89.50@5.03e-3
& 89.00@5.03e-3
& 88.00@3.77e-3
& 85.33@5.57e-4
& 84.17@5.57e-4
& 82.67@5.57e-4 \\
\midrule

\multirow{3}{*}{Lamp}
& ResNet18
& 99.50@0.00
& 99.50@0.00
& 99.50@0.00
& 99.12@6.28e-4
& 99.12@6.28e-4
& 98.75@0.00
& 99.42@0.00
& 99.33@0.00
& 99.33@0.00 \\

& ResNet34
& 99.50@0.00
& 99.50@0.00
& 99.25@0.00
& 98.75@0.00
& 98.25@0.00
& 98.00@0.00
& 99.17@0.00
& 99.00@0.00
& 98.83@0.00 \\

& DeepIrisNet
& 99.00@0.00
& 98.50@0.00
& 98.25@0.00
& 99.00@1.26e-3
& 99.00@1.26e-3
& 99.00@1.26e-3
& 99.17@0.00
& 99.17@0.00
& 99.00@0.00 \\
\bottomrule
\end{tabular}
\end{table*}

\subsection{Main Protected Verification Performance}

We evaluate MPLCE on VGGFace2, FaceScrub, Thousand, and Lamp under enrolled identity numbers \(N\in\{100,200,300\}\) and numbers of permuted-label classifiers \(m\in\{12,16,20\}\). For VGGFace2 and FaceScrub, 40 images are randomly selected per identity, with 20 used for enrollment and 20 for queries. For Thousand, 8 of 10 selected images per identity are used for enrollment and 2 for queries, while Lamp uses 16 enrollment and 4 query images from 20 selected images per identity. Within each dataset and \(N\) setting, all backbone models use the same enrolled identities and sample splits.

Table~\ref{tab:main_performance} reports the protected verification performance under different datasets, backbone models, \(N\), and \(m\). MPLCE achieves high GARs with low FARs on both face and iris modalities. On VGGFace2, GAR remains above 88\% across all evaluated settings, and IR101 achieves a GAR of 90.53\% at a FAR of 3.34\(\times\)10\textsuperscript{-4}\% when \(N=300\) and \(m=20\). On Lamp, ResNet18 achieves a GAR of 99.50\% at a FAR of 0.00\% for all evaluated values of \(m\) when \(N=100\), and maintains 99.33\% GAR at 0.00\% FAR when \(N=300\) and \(m=20\). Performance differences across datasets and backbones reflect the stability of the underlying identity predictions. VGGFace2 generally yields higher GAR than FaceScrub, while Lamp outperforms Thousand under comparable settings. Because a genuine query is accepted only when its predicted label sequence matches the enrolled sequence, reduced prediction stability directly lowers GAR.

Both \(N\) and \(m\) affect protected verification performance. Increasing \(N\) generally increases the difficulty of identity classification and may reduce prediction stability. For example, on VGGFace2 with IR101 and \(m=12\), GAR decreases from 94.40\% to 91.72\% as \(N\) increases from 100 to 300. Increasing \(m\) requires genuine samples to produce consistent predictions across more classifier positions. On FaceScrub with IR50 and \(N=200\), GAR decreases from 85.70\% to 83.47\% as \(m\) increases from 12 to 20. At the same time, a larger \(m\) expands the effective candidate space and can reduce cross-identity template collisions. Therefore, \(m\) introduces a tradeoff between candidate-space expansion and genuine-template consistency.

\subsection{Design Analysis of Multiple Classifiers and Label Permutation}

MPLCE combines multiple classifiers with classifier-specific label permutations. To examine their distinct roles, we compare MPLCE with a single-classifier variant, denoted as Single, and a multi-classifier variant without label permutation, denoted as Multi w/o perm. The variants are evaluated in terms of protected verification performance and the effective offline candidate space of intermediate templates under the defined threat model. We quantify the candidate-space size as
\begin{equation}
B_T=\log_2|\Omega_T|,
\end{equation}
where \(\Omega_T\) denotes the candidate intermediate templates considered in offline enumeration. This quantity is used as a candidate-space diagnostic rather than a complete cryptographic security level.

Single uses one classifier trained with the original labels and therefore has \(N\) candidate label encodings. Multi w/o perm. uses \(m\) independently trained classifiers with the same original label mapping; consequently, each stable label sequence consists of repeated encodings of one original label, and its candidate space also contains \(N\) sequences. In MPLCE, the concrete classifier-specific label permutations are not contained in the compromised protected records, so offline enumeration must consider \(N\) possible labels at each of the \(m\) classifier positions. Accordingly,
\begin{equation}
B_T=
\begin{cases}
\log_2 N, & \text{Single},\\
\log_2 N, & \text{Multi w/o perm.},\\
m\log_2 N, & \text{MPLCE}.
\end{cases}
\end{equation}

We evaluate the three variants using VGGFace2 with IR101 and Lamp with ResNet18. The enrolled identity number is fixed to \(N=200\). Single uses one classifier, whereas both multi-classifier variants use \(m=16\) independently trained classifiers. All variants use the same enrolled identities and sample splits within each setting. Table~\ref{tab:design_analysis} reports \(B_T\), GAR, and FAR. Values in parentheses denote the absolute change in \(B_T\) and the relative changes in GAR and FAR with respect to Single.

\begin{table}[t]
\centering
\caption{Design analysis of multiple classifiers and label permutation under \(N=200\). The multi-classifier variants use \(m=16\). Values in parentheses denote the absolute change in \(B_T\) and the relative changes in GAR and FAR with respect to the Single variant.}
\label{tab:design_analysis}
\scriptsize
\renewcommand{\arraystretch}{1.15}
\setlength{\tabcolsep}{2.6pt}
\begin{tabular*}{\columnwidth}{@{\extracolsep{\fill}}lccc@{}}
\toprule
\textbf{Variant}
&  \textbf{\(B_T\)} \textbf{(bits)}
& \textbf{GAR (\%)}
& \textbf{FAR (\%)} \\
\midrule

\multicolumn{4}{c}{\textbf{Face: VGGFace2 + IR101}} \\
\midrule
Single
& 7.64
& 95.79
& 2.11\(\times\)10\textsuperscript{-2} \\

Multi w/o perm.
& 7.64
& 91.68 (-4.29\%)
& 7.54\(\times\)10\textsuperscript{-4} (-96.43\%) \\

MPLCE
& 122.30 (+114.66)
& 91.92 (-4.04\%)
& 5.03\(\times\)10\textsuperscript{-4} (-97.62\%) \\
\midrule

\multicolumn{4}{c}{\textbf{Iris: Lamp + ResNet18}} \\
\midrule
Single
& 7.64
& 99.17
& 4.16\(\times\)10\textsuperscript{-3} \\

Multi w/o perm.
& 7.64
& 98.50 (-0.68\%)
& 0.00 (-100.00\%) \\

MPLCE
& 122.30 (+114.66)
& 99.12 (-0.05\%)
& 6.28\(\times\)10\textsuperscript{-4} (-84.90\%) \\
\bottomrule
\end{tabular*}
\end{table}

Table~\ref{tab:design_analysis} shows that Single achieves the highest GAR but also the highest FAR in both settings. Introducing multiple independently trained classifiers substantially reduces FAR by imposing consistency across multiple classifier positions, although GAR decreases moderately. However, without label permutation, the effective candidate space remains \(N\), corresponding to \(B_T=7.64\) bits. MPLCE achieves verification performance close to Multi w/o perm., while increasing \(B_T\) to 122.30 bits. These results show that multiple classifiers and label permutation play complementary roles: multiple classifiers impose a stricter verification condition, whereas label permutation prevents repeated original-label encodings and expands the effective offline candidate space from \(N\) to \(N^m\).

\subsection{Parameter Sensitivity Analysis}

We analyze the sensitivity of MPLCE to the number of permuted-label classifiers \(m\) and the number of samples per identity \(K\) used for classifier construction and template enrollment. Since the effect of \(N\) has been examined in the main performance evaluation, we fix \(N=200\) and vary \(m\) and \(K\). VGGFace2 with IR101 and Lamp with ResNet18 are used as representative face and iris settings, respectively. For each dataset, the enrolled identities and verification query samples are fixed across all values of \(K\), while the \(K\) construction and enrollment samples are selected from a separate sample pool. Therefore, varying \(K\) changes the number of samples used for classifier construction and template enrollment without changing the verification query set. The classifiers are trained separately for each value of \(K\).

\begin{table}[t]
\centering
\caption{Parameter sensitivity of MPLCE to \(m\) and \(K\) under \(N=200\). Each entry reports GAR@FAR (\%).}
\label{tab:parameter_sensitivity}
\scriptsize
\renewcommand{\arraystretch}{1.10}
\begin{tabular*}{\columnwidth}
{@{\hspace{4pt}}c@{\extracolsep{\fill}}ccc@{\hspace{4pt}}}
\toprule
\multicolumn{4}{c}{\textbf{Face: VGGFace2 + IR101}} \\
\midrule
\(m\)
& \(K=10\)
& \(K=15\)
& \(K=20\) \\
\midrule
8
& 91.63@2.26\(\times\)10\textsuperscript{-3}
& 92.53@1.63\(\times\)10\textsuperscript{-3}
& 93.18@8.79\(\times\)10\textsuperscript{-4} \\

12
& 90.78@1.63\(\times\)10\textsuperscript{-3}
& 91.75@1.38\(\times\)10\textsuperscript{-3}
& 92.40@6.28\(\times\)10\textsuperscript{-4} \\

16
& 90.10@1.51\(\times\)10\textsuperscript{-3}
& 91.13@8.79\(\times\)10\textsuperscript{-4}
& 91.78@2.51\(\times\)10\textsuperscript{-4} \\

20
& 89.58@1.51\(\times\)10\textsuperscript{-3}
& 90.63@7.54\(\times\)10\textsuperscript{-4}
& 91.20@1.26\(\times\)10\textsuperscript{-4} \\
\midrule

\multicolumn{4}{c}{\textbf{Iris: Lamp + ResNet18}} \\
\midrule
\(m\)
& \(K=8\)
& \(K=12\)
& \(K=16\) \\
\midrule
8
& 94.38@1.26\(\times\)10\textsuperscript{-3}
& 98.13@6.28\(\times\)10\textsuperscript{-4}
& 99.13@0.00 \\

12
& 93.63@1.26\(\times\)10\textsuperscript{-3}
& 98.00@6.28\(\times\)10\textsuperscript{-4}
& 99.13@0.00 \\

16
& 93.13@1.26\(\times\)10\textsuperscript{-3}
& 97.50@0.00
& 98.88@0.00 \\

20
& 92.50@1.26\(\times\)10\textsuperscript{-3}
& 97.38@0.00
& 98.63@0.00 \\
\bottomrule
\end{tabular*}
\end{table}

Table~\ref{tab:parameter_sensitivity} shows that increasing \(m\) generally reduces GAR because genuine samples must produce consistent predictions across more classifier positions, while simultaneously expanding the effective candidate space exponentially with \(m\), since the space contains \(N^m\) possible intermediate templates. In the face setting with \(K=20\), GAR decreases from 93.18\% to 91.20\% as \(m\) increases from 8 to 20; in the iris setting with \(K=16\), it decreases from 99.13\% to 98.63\%. By contrast, increasing \(K\) generally improves GAR by providing more samples for classifier training and more observations for selecting the enrolled intermediate template. With \(m=16\), GAR increases from 90.10\% to 91.78\% in the face setting as \(K\) increases from 10 to 20, and from 93.13\% to 98.88\% in the iris setting as \(K\) increases from 8 to 16. FAR remains below 2.26\(\times\)10\textsuperscript{-3}\% across all settings, with several iris configurations achieving 0.00\%. These results indicate that \(m\) should balance candidate-space expansion against genuine matching performance, whereas a sufficiently large \(K\) improves intermediate-template consistency.

\begin{table*}[t]
\centering
\begin{threeparttable}
\caption{Comparison with representative face and iris BTP methods on benchmark datasets.}
\label{tab:external_comparison}
\scriptsize
\renewcommand{\arraystretch}{1.10}
\setlength{\tabcolsep}{3.0pt}

\begin{tabular*}{0.9\textwidth}{
@{\extracolsep{\fill}}
lccc
@{\hspace{7mm}\extracolsep{\fill}}
lccc
@{}
}
\toprule
\multicolumn{4}{c}{\textbf{Iris}}
&
\multicolumn{4}{c}{\textbf{Face}} \\
\cmidrule(lr){1-4}
\cmidrule(lr){5-8}

\textbf{Method}
& \textbf{Lamp}
& \textbf{Thousand}
& \textbf{E/R/H}
&
\textbf{Method}
& \textbf{YTF}
& \textbf{Yale-B}
& \textbf{E/R/H} \\
\midrule

Zhao et al.~\cite{ref53}
& 97.01@0.01
& 88.95@0.01
& \(\checkmark/\checkmark/\times\)
&
Pandey et al.~\cite{ref32}
& --
& 96.74@0.00
& \(\checkmark/\times/\checkmark\) \\

Lee et al.~\cite{ref15}
& --
& 84.68@0.01
& \(\checkmark/\checkmark/\times\)
&
Chen et al.~\cite{ref33}
& --
& 99.16@1.00
& \(\times/\times/\checkmark\) \\

Zhao et al.~\cite{ref7}
& 79.07@0.01
& --
& \(\checkmark/\checkmark/\times\)
&
Jang et al.~\cite{ref57}
& 97.04@0.10
& --
& \(\checkmark/\checkmark/\times\) \\

Othman et al.~\cite{ref54}
& 79.95@0.01
& 78.50@0.01
& \(\checkmark/\times/\times\)
&
Pinto et al.~\cite{ref58}
& 84.96@0.10
& --
& \(\checkmark/\checkmark/\times\) \\

BioDeepHash~\cite{ref_biodeephash}
& 97.17@0.00
& 92.46@0.00
& \(\checkmark/\checkmark/\checkmark\)
&
BioDeepHash~\cite{ref_biodeephash}
& 95.88@2.02\(\times\)10\textsuperscript{-4}
& \textbf{99.31@0.00}
& \(\checkmark/\checkmark/\checkmark\) \\

MPLCE (Ours)
& \textbf{99.10@0.00}
& \textbf{93.23@5.10\(\times\)10\textsuperscript{-3}}
& \(\checkmark/\checkmark/\checkmark\)
&
MPLCE (Ours)
& \textbf{98.61@5.51\(\times\)10\textsuperscript{-5}}
& 98.92@0.00
& \(\checkmark/\checkmark/\checkmark\) \\

\bottomrule
\end{tabular*}

\begin{tablenotes}[flushleft]
\footnotesize
\item Entries are GAR@FAR values in percent. E/R/H indicate ECC-free,
revocable, and cryptographic-hash-based template storage, respectively.
``--'' indicates that the corresponding result was not reported.
\end{tablenotes}
\end{threeparttable}
\end{table*}

\subsection{Comparison with Existing Methods}

We compare MPLCE with representative BTP methods on commonly used face and iris benchmarks. Because existing studies use different datasets, splits, model configurations, and reporting protocols, not all results are obtained under fully controlled settings. We therefore report published results under their original protocols and evaluate MPLCE using the benchmark setting adopted by BioDeepHash~\cite{ref_biodeephash}. All available identities under each benchmark protocol and \(m=16\) classifiers are used. For Lamp and Thousand, the previously described iris class definitions and query splits are retained, while the remaining images are used for classifier construction and template enrollment. For YTF, 40 images per identity are used for classifier construction and enrollment and 5 for queries. For cropped Yale-B, 10 images per identity are used for classifier construction and enrollment, with the remaining images used as queries following the illumination normalization procedure of BioDeepHash. Table~\ref{tab:external_comparison} reports GAR@FAR and several template protection properties.

MPLCE achieves competitive protected verification performance across the evaluated benchmarks. It obtains 99.10\% GAR at 0.00\% FAR on Lamp, exceeding the BioDeepHash result under the same FAR, and achieves 98.61\% GAR on YTF with a substantially low FAR. On Thousand, MPLCE obtains slightly higher GAR than BioDeepHash, although the reported FARs differ. On Yale-B, MPLCE achieves 98.92\% GAR at 0.00\% FAR, slightly below the 99.31\% reported by BioDeepHash. These results demonstrate competitive performance across both face and iris benchmarks, while differences in protocols should be considered when comparing methods reported under their original settings.

From a security perspective, MPLCE avoids retaining similarity-preserving protected representations and does not rely on error correction codes or biometric-dependent helper data. By storing only cryptographic hash outputs and applying application-specific XOR randomization, it reduces exposure to similarity-based and helper-data-related attacks while supporting revocability and unlinkability across applications.

\section{Conclusion}
In this paper, we proposed MPLCE, a biometric template protection framework for secure biometric verification. MPLCE uses multiple permuted-label classifiers to convert biometric samples into identity-level predicted labels and constructs intermediate templates from the resulting classification outputs. The intermediate templates are then protected by application-specific XOR randomization and cryptographic hashing, while their consistency across genuine samples enables exact-match verification. By using identity classification as the basis for intermediate template generation, MPLCE provides a modality-agnostic template construction principle at the protection level without relying on ECC or biometric-dependent helper data. Under the defined full-disclosure model, the database stores only protected templates and the application-specific XOR string, without retaining intermediate templates, hash inputs, or biometric representations. Security analyses and representative attack evaluations support the irreversibility, revocability, and unlinkability of MPLCE under the considered threat model. Experiments on face and iris datasets, together with analyses of design variants and parameter sensitivity, demonstrate practical protected verification performance with low false acceptance rates and characterize the effects of the main design parameters. Future work will investigate the integration of AI model security techniques to strengthen the classifier stage under classifier-side or system-level compromise scenarios, as well as incremental learning techniques for scalable enrollment of new users while preserving the stability of existing users' intermediate templates.


 

\bibliographystyle{IEEEtran}
\bibliography{ref_clean}

\vspace{11pt}

\end{document}